\documentclass{elsart}
\usepackage{graphicx}
\graphicspath{{./img/}}
\usepackage{cite}
\usepackage{amssymb}
\usepackage{amsmath}
\begin{document}

\begin{frontmatter}

\title{Coherence control for qubits%
\protect\footnote{This work is dedicated to Uli Weiss on the occasion
of his 60th birthday.}}
\author{Karen M.\ Fonseca-Romero},\protect\footnote{On leave of absence from
Universidad Nacional de Colombia.}
\author{Sigmund Kohler}, and
\author{Peter H\"anggi}
\address{Institut f\"ur Physik, Universit\"at Augsburg,
Universit\"atsstra\ss e~1, D-86135 Augsburg, Germany}

\begin{abstract}
We study the influence of an external driving field on the coherence
properties of a qubit under the influence of bit-flip noise.
In the presence of driving, two paradigmatic cases are considered:
(i) a field that results for a suitable choice of the parameters in
so-called coherent destruction of tunneling and (ii) one that commutes
with the static qubit Hamiltonian.
In each case, we give for high-frequency driving a lower bound for the
coherence time.  This reveals the conditions under which the external
fields can be used for coherence stabilization.
\end{abstract}

\end{frontmatter}

% main text
%----------------------------------------------------------------------
\section{Introduction}
\label{sec:introduction}

The experimental realization of one-qubit gates in solid state setups
\cite{Vion2002a,Chiorescu2003a} and two-qubit gates in ion traps
\cite{Leibfried2003a,Schmidt-Kaler2003a} and Josephson junctions
\cite{Pashkin2003a} has demonstrated that these systems provide remarkable
coherence properties although the goal of $10^{-5}$ errors per gate
operation \cite{Steane1998a} has not yet been achieved experimentally.
Theoretical studies of decoherence of two-level systems
\cite{Leggett1987a} have been extended to gate operations in the presence of
an environment in
Refs.~\cite{Loss1998a,Grifoni1999a,Thorwart2000a,Governale2001a,Storcz2003a}.
The unavoidable coupling to external degrees of freedom and the thereby
caused decoherence still presents a main obstacle for the realization of a
quantum computer.  Thus, proposals for the stabilization of the coherence
of qubits are particularly welcome.

A variety of suggestions in this direction relies on the control of
coherence by the influence of external fields.
In particular, it has been proposed to use the physics of the so-called
coherent destruction of tunneling (CDT) for that purpose.
CDT has originally been discovered in the context of tunneling in a driven
bistable potential \cite{Grossmann1991a,Grossmann1991b}.  There, it has been found that a
particle which is initially in the, say, left well of a symmetric bistable
potential, can be prevented from tunneling by the purely coherent
influence of an oscillating driving field.  This effect is stable
against dissipation in the sense that the ac field also decelerates the
dissipative transitions from the left to the right well
\cite{Dittrich1993a,Grifoni1995a,Hartmann2000a} thereby stabilizing
coherence.  For a preparation in a delocalized superposition, it has been
found \cite{Thorwart2000a} that the coherent dynamics is not suppressed.

A different proposal for coherence stabilization is the application of
a sequence of $\pi$-pulses that flip the sign of the qubit-bath coupling
operator resulting in a so-called dynamical decoupling (DD) of qubit and bath
\cite{Viola1998a,Viola1999a}.  A drawback of this scheme is the fact that it
eliminates only noise sources with a frequency below the repetition rates
of the pulses.  This clearly causes practical limitations.
However, these limitations may be circumvented by using a related scheme
 based on continuous-wave driving, i.e.\ one with a harmonic
time-dependence, which allows higher driving frequencies.

In this paper we investigate the coherence properties of a qubit coupled to
an ohmic environment which we model by a spin-boson Hamiltonian
\cite{Weiss1993a}. We consider two different types of harmonic driving:
First, one that leads to coherent destruction of tunneling and, second, one
that corresponds to a continuous wave version of the dynamical decoupling
pulses.  In Section \ref{sec:undriven} we review the spin-boson model for weak
dissipation and derive in Section \ref{sec:DandD} a Markovian formalism for
driven dissipative qubits in the form of a basis-free master equation.  The
global decoherence will be qualified by an upper limit for the decoherence
rate that already gives a reliable estimate for the actual value.
Subsequently, in Sections \ref{CDT} and \ref{DD}, we study the coherence
control by oscillating external fields.

%----------------------------------------------------------------------
\section{Qubit with bit-flip noise}
\label{sec:undriven}

The entire system of the qubit and the environmental degrees of freedom is
described by the microscopic Hamiltonian
\begin{equation}
H = H_\mathrm{qb} + H_\mathrm{coupl} + H_\mathrm{bath} ,
\end{equation}
where the qubit is formed by a two-state system with a level splitting
$\hbar\Delta$, thus
\begin{equation}
\label{eq:H0}
H_\mathrm{qb} = \frac{\hbar\Delta}{2}\sigma_z .
\end{equation}
Below, we will in addition consider an oscillating driving field acting on
the qubit.  The environment is modeled by a bath of harmonic oscillators
which couple linearly to the qubit,
\begin{align}
\label{eq:Hbath}
H_\mathrm{bath} &= \sum_i \hbar\omega_i a_i^\dagger a_i . \\
\label{eq:Hcoupl}
H_\mathrm{coupl} &= \frac{1}{2}\sigma_x\sum_i \hbar c_i (a_i^\dagger+a_i) ,
\end{align}
where $\omega_i$ are the oscillator frequencies and $\hbar c_i$ the
qubit-bath coupling energies.
The bath couples to the qubit operator $\sigma_x$ thereby inducing bit flips,
i.e.\ incoherent transitions between the ground state and the excited state
of the qubit.

We complete the model by choosing an initial condition $\rho_\mathrm{tot}(t_0)$, of
the Feynman-Vernon type, i.e., we assume that at time $t=t_0$ the bath is
in thermal equilibrium and uncorrelated with the qubit, 
\begin{equation}
\label{FVinitial}
\rho_\mathrm{tot}(t_0) = \rho(t_0)\otimes R_\mathrm{bath,eq},
\end{equation}
where $\rho$ is the reduced density operator of the qubit and
$R_\mathrm{bath,eq}\propto\exp(-\beta H_\mathrm{bath})$ is the canonical
ensemble of the bath at the inverse temperature $\beta=1/k_BT$.

\subsection{Dissipative qubit dynamics}

From the Liouville-von Neumann equation $i \hbar \dot
\rho_\mathrm{tot}=[H,\rho_\mathrm{tot}]$ for the total density
operator one obtains for the reduced density operator of the
qubit by standard techniques, the Markovian weak-coupling master equation
\begin{equation}
\begin{split}
\label{eq:Born-Markov}
\dot\rho = -\frac{\mathrm{i}}{\hbar}[H_\mathrm{qb},\rho]
-\frac{1}{4}\int_0^\infty d\tau \Big(
&\mathcal{S}(\tau)[\sigma_x, [\tilde\sigma_x(t-\tau,t), \rho]]
\\
+& \mathcal{A}(\tau)[\sigma_x, \{\tilde\sigma_x(t-\tau,t), \rho\}]\Big) ,
\end{split}
\end{equation}
where $\{A,B\}=AB+BA$ stands for the anticommutator.  The notation $\tilde
X(t,t')$ is a shorthand for $U^\dagger(t,t') X U(t,t')$ with $U$ being the
propagator for the coherent qubit dynamics.  Note that for later use, we
have written the master equation already in a form which is also valid in
the case of an explicit time-dependence of the qubit Hamiltonian.  The
dissipative equation of motion \eqref{eq:Born-Markov} depends on the bath
through the symmetric and anti-symmetric correlation functions of the bath
operator $\mathcal{B}(t)= \sum_i c_i ({a}_i^\dagger \exp(i\omega_i t)+{a}_i
\exp(-i\omega_i t))$, given by
\begin{align}
\label{eq:S}
  \mathcal{S}(\tau)
  &=\frac{1}{2}\left\langle
  \left\{\mathcal{B}(\tau),\mathcal{B}(0)\right\}\right\rangle =
  \frac{1}{\pi}\int_0^\infty \d\omega J(\omega) 
  \coth(\hbar\omega\beta/2)\cos(\omega \tau),
\\
\label{eq:A}
  \mathcal{A}(\tau)
  &=\frac{1}{2}\left\langle
  \left[\mathcal{B}(\tau),\mathcal{B}(0)\right]\right\rangle =
  -\frac{\mathrm{i}}{\pi}\int_0^\infty \d\omega J(\omega) \sin(\omega \tau),
\end{align}
respectively. The angular brackets $\langle \cdots \rangle$ denote the
average with respect to the thermal equilibrium of the bath.
We have introduced for the qubit-bath coupling the spectral density
\begin{equation}
  \label{eq:spectraldensity}
  J(\omega) = \pi\sum_i c_i^2  \delta (\omega-\omega_i).
\end{equation}
If the bath modes are dense, $J(\omega)$ becomes a smooth function.
Within the present work, we will consider the case of Ohmic dissipation where
$J(\omega)= 2\pi\alpha\omega \exp(-\omega/\omega_c)$ and
$\alpha$ is a dimensionless measure for the dissipation strength.
The so-called cutoff frequency $\omega_c$ is assumed to be the highest
frequency of the bath.
If the cutoff frequency is much larger than all relevant energy scales, the
antisymmetric correlation function $A(\tau)$ can be replaced by
$2\pi\mathrm{i}\alpha\delta'(\tau)$.
Then, for the integral in the master equation \eqref{eq:Born-Markov}, the
part containing $A(\tau)$ can be evaluated after integrating by parts
to read
\begin{equation}
\label{eq:MEpart2}
\frac{\pi\alpha}{4}[\sigma_x,\{[H_\mathrm{qb}/\hbar,\sigma_x],\rho\}].
\end{equation}

To bring the master equation \eqref{eq:Born-Markov} into a more explicit
form, we insert the Heisenberg operator
\begin{equation}
\tilde\sigma_x(t-\tau,t)=\sigma_x\cos(\Delta\tau)+\sigma_y\sin(\Delta\tau),
\end{equation}
which is readily derived from its definition together with the qubit
Hamiltonian \eqref{eq:H0}.  Performing the integration over $\tau$ and
$\omega$ and neglecting renormalization effects, which are small provided
that $\alpha\ln (\omega_c/\Delta)\ll 1$, yields for $\Delta\ll\omega_c$,
the Markovian master equation
\begin{equation}
\label{eq:ME1}
\dot\rho = -\frac{\mathrm{i}}{\hbar}[H_\mathrm{qb},\rho]
-\frac{\Gamma}{4}[\sigma_x,[\sigma_x,\rho]]
+\mathrm{i}\frac{\pi\alpha\Delta}{4}[\sigma_x,\{\sigma_y,\rho\}].
\end{equation}
We will find below that the relaxation and decoherence processes are
determined by the rate \cite{Weiss1993a}
\begin{equation}
  \Gamma = \frac{1}{2}\mathcal{S}(\Delta) ,
\end{equation}
where
\begin{equation}
  \label{eq:S.omega}
  \mathcal{S}(\omega) = 2\pi\alpha\omega%\exp(-\omega/\omega_c)
  \coth\left(\frac{\hbar\omega\beta}{2}\right)
\end{equation}
is the power spectrum of the bath fluctuations, i.e.\ the Fourier
transformed
of the symmetric bath correlation function \eqref{eq:S}.  This master
equation is a basis-free version of a Bloch-Redfield or Floquet-Markov master
equation \cite{Blumel1991a,Kohler1997a}.
Such an operator notation is preferable to a decomposition into the qubit's
eigenbasis for ease of notation and, moreover, since it allows a more
elegant computation of expectation values.  The interpretation of
\eqref{eq:ME1} is that the first term of the right-hand side is responsible
for the coherent dynamics, while the second and third term correspond to
decoherence and relaxation.

For the description of dynamics of a single qubit, it is convenient to map
the density operator to the Bloch vector $\vec{s} = \mathop{\mathrm{tr}}
(\vec \sigma \rho)$.  It is straightforward to derive for the Bloch vector
from the master equation \eqref{eq:ME1} the inhomogeneous linear equation
of motion
\begin{equation}
\label{eq:bloch}
  \dot{\vec s} = -M \vec s + \vec b ,
\end{equation}
where
\begin{equation}
M =
\left(\begin{array}{ccc} 
0 & -\Delta & 0 \\ 
\Delta & \Gamma & 0 \\  
0 & 0 & \Gamma
\end{array} \right),
\quad
\vec b =\left(\begin{array}{c} 0\\ 0\\ -\pi\alpha\Delta \end{array}\right).
\end{equation}
In the weak dissipation limit, $\Gamma\ll \Delta$, the matrix $M$ has the eigenvalues
\begin{equation}
\label{eq:eigenvalues}
  \Gamma, \quad
  \frac{1}{2}\Gamma \pm \mathrm{i} \Delta .
\end{equation}
While the first eigenvalue describes the decay of the population of the
excited state, the other two correspond to damped oscillations
of the off-diagonal density matrix elements in the basis eigenbasis of the
qubit Hamiltonian.
This justifies for $\Gamma$ the designation \textit{relaxation rate}
and the definition of the \textit{decoherence rate} $\Gamma_\phi=\Gamma/2$.
For larger systems there is room for several relaxation (purely real
eigenvalues) and decoherence rates (complex eigenvalues with non-vanishing
imaginary parts).

\subsection{Entropy production}

Relaxation and decoherence are nonunitary processes, i.e.\ processes
which transform a pure state into a mixed state.  Since a quantum computer
relies fundamentally on the coherence of the time evolution, such
processes put an essential limitation and their influence has to be
minimized.  Thus, in order to follow the process of coherence loss,
it is desirable to define proper coherence measures which preferably
are independent of the chosen basis.
A natural possibility that comes to mind is the Shannon entropy
$\mathop{\mathrm{tr}}(\rho\ln\rho)$.  However, this measure is sometimes
inconvenient due to the appearance of the logarithm.  Therefore, it is
common to use instead the ``linear entropy''
\cite{Watanabe1939a,Nemes1986a,Zurek1993a}
\begin{equation}
S = 1-\mathop{\mathrm{tr}}(\rho^2)
=\frac{1}{2}(1-\vec s\cdot\vec s),
\end{equation}
which is a good approximation for the Shannon entropy for almost pure
states and is closely related to the so-called purity
$\mathop{\mathrm{tr}}(\rho^2)$.  The history of this kind of measure is
long, and can be traced back to at least 1939 \cite{Watanabe1939a}.
The linear entropy possesses a convenient physical interpretation:
Suppose that $\rho$ describes an incoherent mixture of $n$ orthogonal
states with equal probability, then $\mathop{\mathrm{tr}}(\rho^2)$ reads
$1/n$. For a single qubit the maximum linear entropy is therefore 1/2.  It
is zero if and only if $\rho$ describes a pure state.

The decoherence rate is well described by the entropy production
\begin{equation}
\dot S = -2\mathop{\mathrm{tr}}(\rho\dot\rho)
= - \vec s \cdot \dot{\vec s}
\end{equation}
which follows directly from the master equation, respectively from the
equation of motion for the Bloch vector.  Obviously, the coherent dynamics
given by the first commutator in the master equation \eqref{eq:ME1} does
not increase the entropy.

In the context of quantum computing, we are mainly interested in the entropy
production for (almost) pure states, i.e.\ the initial entropy production
which is determined by the eigenvalues \eqref{eq:eigenvalues} of the matrix
$M$.  Since these eigenvalues are invariant under unitary transformations,
they represent a global measure for the influence of the dissipation.
In particular, they are independent of the choice of the basis.
In the worst case, the decoherence is determined by the eigenvalue with the
largest real part.

For the driven system considered below, the matrix becomes time-dependent
and, consequently, the eigenvalues \eqref{eq:eigenvalues} must be replaced
by the eigenvalues of a Floquet equation.  Their computation, however,
might be quite cumbersome.  Therefore, a more convenient measure which is
also applicable to driven systems is the sum $\gamma$ of all eigenvalues
which can be computed without explicit diagonalization by the trace of the
matrix $M$,
\begin{equation}
\label{eq:trace}
\gamma \equiv \mathop{\mathrm{tr}}M = 2\Gamma .
\end{equation}
Since $\gamma$ is real and larger than all real parts of the eigenvalues
(which are positive), it represents an upper bound for the decoherence
rate.  Moreover, for the present case, the largest real part of an
eigenvalue is at least $\gamma/3$ and, thus, we can conclude that $\gamma$
gives the correct order of magnitude for the decoherence rate of the most
fragile initial state.

A further related, but more probabilistic measure, that characterizes the
loss of coherence, is the average $\dot S$ over all pure initial states of
the entropy production and reads
\begin{equation}
\Gamma_{\mathrm{av}} \equiv \langle\dot{S}(0)\rangle_{\mathrm{av}}
= \frac{\gamma}{3} .
\end{equation}

%----------------------------------------------------------------------
\section{Dissipation and driving}
\label{sec:DandD}

To study the controllability of the decoherence, we generalize
the master equation to the case where a time-dependent Hamiltonian
\begin{equation}
H_D(t) = H_1 f(t)
\end{equation}
acts on the qubit.  We assume that time-dependence $f(t)$ is periodic
with the driving period $T=2\pi/\Omega$ and has zero average.
Then, according to the Floquet theorem \cite{Manakov1986a}, the
corresponding propagator can always be written in the form
\begin{equation}
\label{eq:UFgeneral}
U(t,t') = U_P(t,t') U_F(t-t')
\end{equation}
where $U_P(t,t')=U_P(t+T,t')$ obeys the time-periodicity of the driving
field.  The Floquet propagator $U_F$ depends only on the time difference
and contains non-adiabatic phases which emerge during the
propagation over one driving period \cite{Aharonov1987a}.
Since $U_P$ is time-periodic, the long-time dynamics of the driven qubit
is entirely determined by the one-period propagator $U_F(T)$.

Inserting the propagator \eqref{eq:UFgeneral} into the master equation
\eqref{eq:ME1} results after the $\tau$-integration in a dissipative kernel
that still depends periodically on the final time $t$.  However, since we
consider the case of fast driving, it is possible to separate time-scales
and replace the rapidly oscillating operators in the master equation by
their time-averages.  In doing so, we arrive at
\cite{Kohler1997a,Hartmann2000a}
\begin{equation}
\label{eq:MEF}
\dot\rho = -\frac{\mathrm{i}}{\hbar}[H_\mathrm{qb}+H_D(t),\rho]
-\frac{1}{4}[\sigma_x,[Q,\rho]]
+\mathrm{i}\frac{\pi\alpha\Delta}{4}[\sigma_x,\{\sigma_y,\rho\}],
\end{equation}
where the difference with respect to the master equation \eqref{eq:ME1} for
the static case comes from replacing in the second commutator
$\Gamma\sigma_x$ by the operator
\begin{equation}
\label{eq:OperatorQ}
Q = \frac{1}{T}\int_0^T\mathrm{d} t \int_0^\infty \mathrm{d}\tau\,
\mathcal{S}(\tau)U_F^\dagger(\tau)
U_P^\dagger(t-\tau,t) \sigma_x
U_P(t-\tau,t) U_F(\tau) .
\end{equation}
The master equation \eqref{eq:MEF} for the driven, dissipative qubit
reflects the close resemblance to the master equation \eqref{eq:ME1} valid
in the static case.  The last term in \eqref{eq:MEF} is not modified by the
driving.  The reason for this is that the driving enters linearly in
\eqref{eq:MEpart2} and, consequently, vanishes in the average over the
driving period.  However, besides the explicit presence of the
time-dependent Hamiltonian in the coherent contribution, also the
dissipative part has acquired a change: The coupling operator $\sigma_x$ is
now replaced by an operator $Q$ that depends on the qubit propagator in the
presence of the driving.  As a consequence, we expect in the case of strong
driving fields not only a modification of the coherent dynamics, but also
of the coherence properties.  In the following sections, we will
investigate two typical types of driving: as a result, we find that they
can alter the coherence times rather significantly.

%----------------------------------------------------------------------
\section{Coherent destruction of tunneling}
\label{CDT}

As a first example for the significant influence of the driving field, we
investigate the qubit under the influence of the Hamiltonian
\begin{equation}
\label{eq:HF:CDT}
H_D(t) = A\sigma_x\cos(\Omega t)
\end{equation}
which couples like the bath to the qubit by via the operator $\sigma_x$
and, thus, commutes with the qubit-bath coupling.  Such a time-dependent
field causes already interesting effects for the coherent qubit dynamics
that we will briefly review before discussing decoherence.

To sketch the coherent dynamics of the driven qubit, we first transform the
Hamiltonian $H_\mathrm{qb}+H_D(t)$ by the unitary operator
\begin{equation}
\label{eq:Ipicture}
U_0(t) =
\exp\left(-\mathrm{i}\frac{A\sigma_x}{\hbar\Omega}
          \sin(\Omega t)\right) .
\end{equation}
This transformation defines the interaction picture with respect to
$H_D(t)$ and results in the likewise $T$-periodic Hamiltonian
\begin{equation}
\label{eq:Hinteraction}
\tilde H_\mathrm{qb}(t)
= \frac{\hbar\Delta}{2}\left\{
   \sigma_z\cos\left(\frac{2A}{\hbar\Omega}\sin(\Omega t)\right)
  +\sigma_y\sin\left(\frac{2A}{\hbar\Omega}\sin(\Omega t)\right)
  \right\} .
\end{equation}
The corresponding Schr\"odinger equation cannot be integrated exactly since
$\tilde H_\mathrm{qb}(t)$ does not commute with itself at different times and, thus,
time-ordering has to be taken into account.
We restrict ourselves to an approximate solution and
neglect corrections of the order $\Delta^2$.  Within this approximation,
the propagator is simply given by the exponential of the integral of the
time-dependent Hamiltonian.  This consists of two parts: First, there is
the time-average of $\tilde H_\mathrm{qb}(t)$ which determines the 
leading contribution to the Floquet propagator $U_F(t)$.
The second part comprises terms that oscillate with the driving period.
For a high-frequency driving with $\Omega\gg\Delta$, this latter
contribution can be neglected since the periodic part of the propagator
is dominated by the contribution $U_0$ coming directly from the
driving.  This approach is equivalent to a perturbational computation of
Floquet states \cite{Shirley1965a,Grifoni1998a} as has been shown in
\cite{Kohler2003b}.

Finally, the qubit propagator in the interaction picture defined by
\eqref{eq:Ipicture}, is determined by the time average
of the Hamiltonian \eqref{eq:Hinteraction} which reads
\begin{equation}
\bar H_\mathrm{qb} = \frac{\hbar\Delta_\mathrm{eff}}{2}\sigma_z .
\end{equation}
This static approximation to the driven qubit Hamiltonian is of the same
form as the original Hamiltonian \eqref{eq:H0}, but with the tunneling
matrix element renormalized according to
\begin{equation}
\label{eq:Delta:eff}
\Delta \to \Delta_\mathrm{eff}=J_0(2A/\hbar\Omega)\Delta .
\end{equation}
Here, $J_0$ is the zeroth order Bessel function of the first kind.
Thus, within this high-frequency approximation, the entire propagator 
for the qubit in the Schr\"odinger picture reads
\begin{equation}
\label{eq:U(t,0)}
  U(t,0)
  = \exp\left(-\mathrm{i}\frac{A}{\hbar\Omega} \sin(\Omega t)\sigma_x \right)
  \exp\left(-\mathrm{i}\frac{\Delta_\mathrm{eff}t}{2}\sigma_z \right) .
\end{equation}

Of particular interest are now driving parameters for which $2A/\hbar\Omega$
corresponds to a zero of the Bessel function $J_0$, i.e.\ for which
$\Delta_\mathrm{eff}$ vanishes.  Then the one-period propagator $U_F(T)$,
becomes the identity.  Or in other words: the long-time dynamics is
suppressed.
The dynamics within the driving period requires a closer look at the
periodic propagator $U_P$: For an initial preparation in an
eigenstate of $\sigma_x$, it contributes only a global
phase, such that the dynamics as a whole is suppressed also within the
driving period.
This effect of suppressing the
time-evolution by the purely coherent influence of an external field has
been investigated first in the context of driven tunneling
\cite{Grossmann1991a,Grossmann1991b} and is named
``coherent destruction of tunneling'' (CDT)\footnote{Although
we consider here also parameters for which the dynamics is not entirely
suppressed, we will refer to a driving of the form \eqref{eq:HF:CDT} as
``CDT driving''.}.  However, for a preparation other than an eigenstate of
$\sigma_x$, the periodic propagator $U_P$ will still cause a non-trivial
dynamics within the driving period.  This fact has been found numerically
in Ref.~\cite{Thorwart2000a}.

There is also a significant influence of the driving \eqref{eq:HF:CDT} on
the dissipative dynamics of a particle with an initial preparation in an
eigenstate of $\sigma_x$: Besides slowing down the coherent time evolution,
also the dissipative transitions are decelerated, i.e. the rate for the
dissipative transitions between the wells of a bistable system becomes
lower.  In the following, we investigate decoherence and dissipation
under such a CDT driving without the restriction to a specific preparation.

In order to evaluate the coefficients of the master equation \eqref{eq:MEF}
valid in the driven case, we have to compute the operator $Q$.  This
requires an explicit expression for the propagator $U(t,t')$ starting at
time $t'$ and, thus, we have to consider initial phases.  Repeating the
calculation from the beginning of this section yields
\begin{equation}
  \label{eq:UFloquet}
  U(t-\tau,t) =  \exp\left(-\mathrm{i}\frac{A}{\hbar\Omega}
  [\sin\Omega(t-\tau)-\sin\Omega t]\sigma_x \right)
  \exp\left(-\mathrm{i}\frac{\Delta_\mathrm{eff}\tau}{2}\sigma_z \right) ,
\end{equation}
where the first factor is the time-periodic part $U_P(t-\tau,t)$ while the
second factor determines the long-time dynamics.
Since $U_P$ commutes with $\sigma_x$, only the Floquet propagator $U_F$
is relevant for the operator \eqref{eq:OperatorQ}.  After performing the time
and the frequency integration we find 
\begin{equation}
\label{eq:Q_CDT}
Q_\mathrm{CDT} = \frac{1}{2}\mathcal{S}(\Delta_\mathrm{eff})\sigma_x .
\end{equation}
Thus, the master equation is of the same form as in the undriven case,
Eq.~\eqref{eq:ME1}, but with $\Gamma$ replaced by
\begin{equation}
\label{eq:GammaCDT}
\Gamma_\mathrm{CDT}=\frac{1}{2}\mathcal{S}(\Delta_\mathrm{eff})
= \Gamma J_0(2A/\hbar\Omega)
\frac{\coth(\beta\hbar\Delta_\mathrm{eff}/2)}{\coth(\beta\hbar\Delta/2)}.
\end{equation}
Since the spectral density $S(\omega)$ increases monotonically with the
frequency $\omega$ and, moreover, the Bessel function $J_0(x)<1$ for $x>0$,
dissipation and decoherence become smaller due to the driving.

Again, we employ the trace of the matrix $M$ appearing in the equation of
motion for the Bloch vector as a criterion for the global decoherence
strength and find
\begin{equation}
\label{eq:gammaCDT}
  \gamma_\mathrm{CDT} = \mathcal{S}(\Delta_\mathrm{eff})
\end{equation}
which represents within the high-frequency approximation an upper limit for
the decoherence rates.
For high temperatures, $\beta\hbar\Delta_\mathrm{eff}\ll 1$, the $\coth$ in
the power spectrum \eqref{eq:S.omega} can be replaced by the inverse of its
argument.  This defines the classical limit where the decoherence rate
\eqref{eq:gammaCDT} becomes $4\pi\alpha k_BT/\hbar$, i.e.\ independent of
the effective tunnel splitting $\Delta_\mathrm{eff}$.
In the low temperature limit $\beta\hbar\Delta_\mathrm{eff}\gg 1$, the
$\coth$ becomes unity and, thus,
\begin{equation}
\label{eq:gammaCDT2}
  \gamma_\mathrm{CDT} \approx \gamma J_0(2A/\hbar\Omega) ,
\end{equation}
which means that decoherence becomes smaller by a factor
$J_0(2A/\hbar\Omega)$.  This reduction of decoherence is brought about by
the fact that the driving \eqref{eq:HF:CDT} decelerates the long time
dynamics of the qubit.  Thereby, the frequencies which are relevant for the
decoherence are shifted to a range where the spectral density of the bath
is lower.  Consequently, the influence of the bath is diminished.  Since
$\gamma_\mathrm{CDT}$ represents an upper bound for the decoherence rate,
this demonstrates that coherence stabilization via CDT is a phenomenon that
is independent of the initial state.

It should be noted, however, that the \textit{coherent} dynamics is slowed
down by the same factor as the \textit{decoherence}, cf.\
Eqs.~\eqref{eq:Delta:eff} and \eqref{eq:gammaCDT2}.  Thus, if for a
specific application, the figure of merit is the number of coherent
oscillations, the present coherence stabilization scheme may not prove
useful.
For resonant driving, this unfavorable situation can change for certain
preparations, see Fig.~2 in Ref.~\cite{Thorwart2000a}.

%----------------------------------------------------------------------
\section{Dynamical decoupling}
\label{DD}

A recently proposed mechanism for coherence stabilization is the so-called
dynamical decoupling \cite{Viola1998a}.  This scheme employs sequences of
$\pi$-pulses which flip the sign of the operator that couples the qubit to
the bath operators.  The basic idea originates from the suppression of spin
diffusion in nuclear magnetic resonance experiments
\cite{Carr1954a,Haeberlen1968a} and
has become an established technique in that area \cite{Slichter1990a}.
In the present case where the bath couples to the operator
$\sigma_x$ [cf.\ Eq.~\eqref{eq:Hcoupl}], such a transformation is
e.g.\ induced by the Hamiltonian $\hbar\omega_R\sigma_z$ for a pulse duration
$\pi/\omega_R$.  Since the corresponding propagator is a function of the
qubit Hamiltonian, the coherent dynamics is not altered.
Besides the prospective benefits of such a control scheme, there is also a
number of possible drawbacks that the application of $\pi$-pulses might
cause: For a driven system, there is always the possibility of unwanted
off-resonant transitions \cite{Tian2000a}, especially in the case of sharp
pulses.  A more practical limitation is the fact that only noise with
frequencies below the pulse repetition rate can be eliminated in such a way.

These disadvantages can be overcome partially by applying a continuous wave
version of the dynamical decoupling scheme, i.e.\ a driving of the form
\begin{equation}
\label{eq:HFDD}
H_D = A\sigma_z \cos(\Omega t)
\end{equation}
for which the available frequency range is larger.
This constitutes our second example of a driving field influencing
considerably the qubit decoherence.
In contrast to the driving Hamiltonian employed in the previous section, the
present one commutes with the static qubit Hamiltonian and,
consequently, the propagator for the driven qubit can be computed exactly
to read
\begin{equation}
U(t,t')
= \exp\left(-\mathrm{i}\frac{A}{\hbar\Omega}[\sin(\Omega t)- \sin(\Omega t')]
  \sigma_z\right)
  \exp\left(-\mathrm{i}\Delta\sigma_z(t-t')\right) .
\end{equation}
Again, we have written the propagator in the form \eqref{eq:UFgeneral}
which is suitable for simplifying the master equation \eqref{eq:MEF}.
Inserting this into the expression \eqref{eq:OperatorQ} results for
$\Delta\ll\Omega$ in the effective coupling operator
\begin{equation}
\label{eq:QDD}
Q_\mathrm{DD} = \frac{1}{2}
\sigma_x\left(J_0^2(2A/\hbar\Omega) \mathcal{S}(\Delta)
+2 \sum_{n=1}^\infty J_n^2\left({2 A}/{\hbar\Omega}\right)
\mathcal{S}(n\Omega) \right) .
\end{equation}
In order to derive this expression, we have decomposed the exponentials of
the trigonometric functions into a Fourier series using the identity
$\exp[\mathrm{i}x\sin(\Omega t)]=\sum_k J_k(x)
\exp(\mathrm{i}k\Omega t)$, where $J_k$ is the $k$th order Bessel function
of the first kind \cite{Gradshteyn1994a}.  Like in the previous cases, the
effective coupling operator $Q_\mathrm{DD}$ is proportional to $\sigma_x$
and, thus, the master equation is again of the form \eqref{eq:ME1}.  The
only change for the dissipative dynamics is the replacement of $\Gamma$ by
\begin{equation}
\label{eq:GammaDD}
\Gamma_{\mathrm{DD}}= \Gamma
\left\{
J_0^2\left(2A/\hbar\Omega\right)
+2 \sum_{n=1}^\infty
\frac{n\Omega}{\Delta}
\frac{\tanh(\hbar\Delta\beta/2)}{\tanh(n\hbar\Omega\beta/2)}
\mathrm{e}^{-n\Omega/\omega_c} J_n^2\left(2A/{\hbar\Omega}\right) \right\}.
\end{equation}
The decoherence rate in this case depends on the spectral density of
the bath at multiples of the driving frequency which may be larger than the
cutoff frequency $\omega_c$.
The $\pi$-pulses applied in the original version version \cite{Viola1998a}
of dynamical decoupling, correspond for a continuous driving to a field
amplitude such that $2A/\hbar\Omega$ equals the first zero of the Bessel
function $J_0$, i.e.\ assumes a value $2.404825\ldots$.
Then only the sum in Eq.~\eqref{eq:GammaDD} contributes to the decoherence 
rate $\Gamma_\mathrm{DD}$.
If now the driving frequency is larger than the cutoff of the spectral
density, $\Omega>\omega_c$, the decoherence rate is considerably reduced:
For low temperatures, $1/\beta\ll\hbar\Delta$, the hyperbolic tangent in the
decoherence rate \eqref{eq:GammaDD} become unity and each contribution is
weighted by a possibly large factor $n\Omega/\Delta$.  In the
high-temperature limit $1/\beta\ll\hbar\Omega$, we use $\tanh x\approx x$
and find that the dependence of the prefactor on
$n\Omega$ cancels.  This means that the dynamical decoupling scheme is
especially useful for high temperatures.  The physical reason for this is
that the driving shifts the qubit dynamics towards high frequencies where
the thermal occupation of the bath modes is negligible.

\begin{figure}[t]
\centering
\includegraphics[width=0.80\columnwidth]{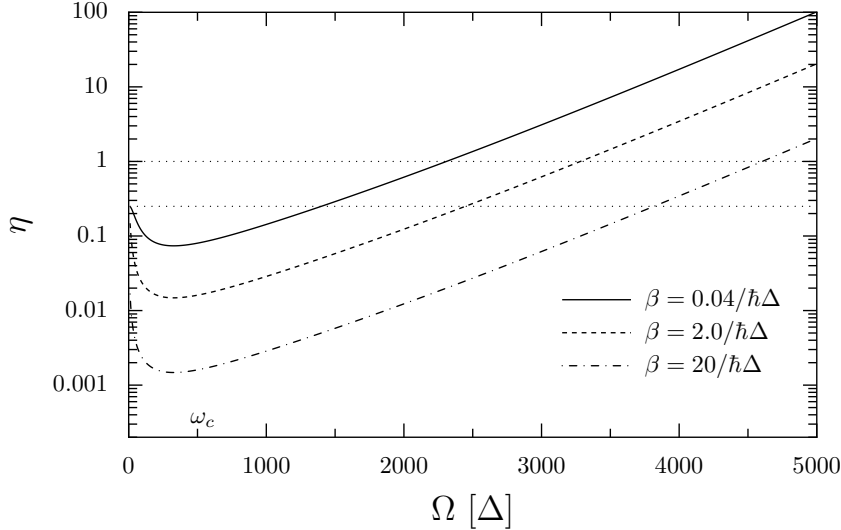}
\caption{Coherence stabilization factor $\eta$ as a function of the driving
frequency for dynamical decoupling for different temperatures.
The cutoff frequency is $\omega_c=500\Delta$, $2A/\hbar\Omega=2.4$,
and the dissipation strength is $\alpha=0.01$.
The dotted horizontal lines mark the values above which the
coherence is improved on average ($\eta>1/4$) and for an arbitrary initial
condition ($\eta>1$), respectively.
\label{fig:DD}}
\end{figure}%
To assess the coherence stabilization originating from dynamical
decoupling, we define for the coherence stabilization the rather
conservative measure
\begin{equation}
\eta = \frac{\Gamma/2}{\gamma_\mathrm{DD}} ,
\end{equation}
i.e.\ the ratio between the lowest decoherence rate \textit{without} driving
and the largest decoherence rate \textit{with} driving, cf.\ equations
\eqref{eq:eigenvalues} and \eqref{eq:trace}.
For parameters such that $\eta>1$, it is granted that the driving
stabilizes the coherence independent of the initial state.  If $\eta>1/4$,
we still can conclude that the coherence is improved in the average over all
possible initial states.

Figure \ref{fig:DD} compares the coherence stabilization $\eta$ as a
function of the driving frequency for $2A/\hbar\Omega=2.4$, i.e.\ close to
a zero of the Bessel function $J_0$.  It reveals that for driving
frequencies well below the cutoff, the driving rather spoils the
coherence.  This improves with increasing driving frequency and, finally,
for a high-frequency driving $\eta$ becomes much larger than unity
corresponding to a significant coherence stabilization.
The data demonstrate the usefulness of dynamical decoupling for high
temperatures discussed above.

%----------------------------------------------------------------------
\section{Conclusions}

We have investigated the coherence properties of a qubit weakly coupled to
a harmonic oscillator bath under the influence of an external driving
field.  Two types of driving have been taken into account: One that can
cause coherent destruction of tunneling and another that corresponds to
$\pi$-pulses that invert the sign of the qubit-bath coupling operator.  To
estimate the decoherence, we have derived an upper bound for the
decoherence rate.  As a result, we have found that both types of driving
can enhance the coherence properties of the qubit significantly,
independent of the preparation.

In the case of CDT we have found that the main effect comes from the fact
that the driving shifts the coherent long-time dynamics of the qubit
towards lower frequencies where the spectral density of an ohmic bath is
lower and, thus, the effective coupling is weaker.  Consequently,
decoherence is generally reduced.  This is most pronounced at low
temperatures.  For high temperatures, however, the lower spectral density
is counterbalanced by an increasing thermal noise, such that decoherence
becomes independent of the driving.

For the continuous-wave dynamical decoupling scheme, we have found that a
low-frequency driving is counterproductive.  However, once the frequency
exceeds the bath cutoff, the coherence properties recover and are finally
significantly improved, especially at high temperatures.  Since such a
dynamical decoupling by a harmonic driving allows higher driving
frequencies than the pulsed version, this form of coherence stabilization
bear interesting perspectives for applications.

%----------------------------------------------------------------------
\section*{Acknowledgements}

This work has been supported by the research network ``Quanteninformation
entlang der A8'' and the Sonderforschungsbereich 631 of the Deutsche
Forschungsgemeinschaft.

%-----------------------------------------------------------------------------

\end{document}